\documentclass[12pt]{iopart}
\usepackage{iopams}
\usepackage{epsfig}
\begin{document}

\title{Hard spectra and QCD matter: experimental review}

\author{David~d'Enterria\dag\
\footnote[3]{e-mail:denterria@nevis.columbia.edu}
}

\address{\dag\ Nevis Laboratories, Columbia University\\ 
Irvington, NY 10533, and New York, NY 10027, USA}

%9 pages, 9 figures, plenary talk at the 17th International Conference 
%on Ultra-Relativistic Nucleus-Nucleus Collisions (Quark Matter 2004), 
%to appear in the Proceedings (Journal of Physics G)

\begin{abstract}
The most significant experimental results on hadron spectra 
at large transverse momentum available at the time of Quark Matter 
2004 conference are reviewed. Emphasis is put on those measurements 
that provide insights on the properties of the QCD media, 
``Quark Gluon Plasma'' and ``Color Glass Condensate'', expected to 
be present in nucleus-nucleus collisions at collider energies.
\end{abstract}


\submitto{\JPG}

%\maketitle

\section{Introduction}

Nucleus-nucleus collisions at relativistic energies aim at the 
study of the fundamental theory of the strong interaction, 
Quantum Chromo Dynamics (QCD), at extreme energy densities. 
The main goal of this physics program is the production and 
study under laboratory conditions of the plasma of quarks and 
gluons (QGP). The QGP is a deconfined and chirally symmetric 
state of strongly interacting matter predicted by QCD calculations 
on the lattice~\cite{latt} for values of the energy density 
five times larger than those found in the nuclear ground state,
$\epsilon\gtrsim$ 0.7 $\pm$ 0.3 GeV/fm$^3$.
The combination of high center-of-mass energies and large nuclear 
systems in the initial-state of heavy-ion reactions provides, furthermore, 
favorable conditions for the study of the (non-linear) parton dynamics 
at small values of (Bjorken) fractional momentum $x$. In this regime
(often  dubbed ``Color Glass Condensate'', CGC~\cite{cgc}), %where 
higher-twist effects are expected to saturate the rapidly increasing 
density of ``wee'' gluons observed at small-$x$ in the hadronic wave 
functions which would, otherwise, violate the unitarity limit of 
the theory.

In hadronic collisions the production of particles with high 
transverse momentum ($p_{T} \gtrsim$ 2 GeV/$c$) results from 
hard parton-parton scattering processes with large momentum 
transfer $Q^2$ and, as such, is directly connected to %provides direct information on 
the fundamental (quark and gluon) degrees of freedom of QCD.
Since hard cross-sections can be theoretically computed by perturbative 
methods using the collinear factorization theorem~\cite{factor},
inclusive high $p_T$ hadroproduction, jets, %Drell-Yan, 
direct photons, and heavy flavors, have long been considered 
sensitive and well calibrated probes of the small-distance 
QCD phenomena. This paper reviews the most interesting 
results on transverse spectra from $Au+Au$ reactions at RHIC collider 
energies ($\sqrt{s_{_{NN}}}$ = 200 GeV) in the high $p_T$ sector, 
where the production of hadrons in central collisions shows substantial 
differences compared to heavy-ion reactions at lower center-of-mass-energies
($\sqrt{s_{_{NN}}}\approx$ 20 GeV) as well as compared to more 
elementary reactions either in the ``vacuum'' ($p+p$, $e^+e^-$) or 
in a cold nuclear matter environment ($d,l+A$). Such differences are 
indicative of significant initial- and final- state effects and %are directly connected to 
provide direct information on the properties of the QCD medium in 
which the hard scattering process has taken place.

\section{Hard spectra in the QCD vacuum: $p+p$ collisions}

\begin{figure}[htbp]
\begin{center}
\begin{minipage}[t]{.46\linewidth}
\epsfig{file=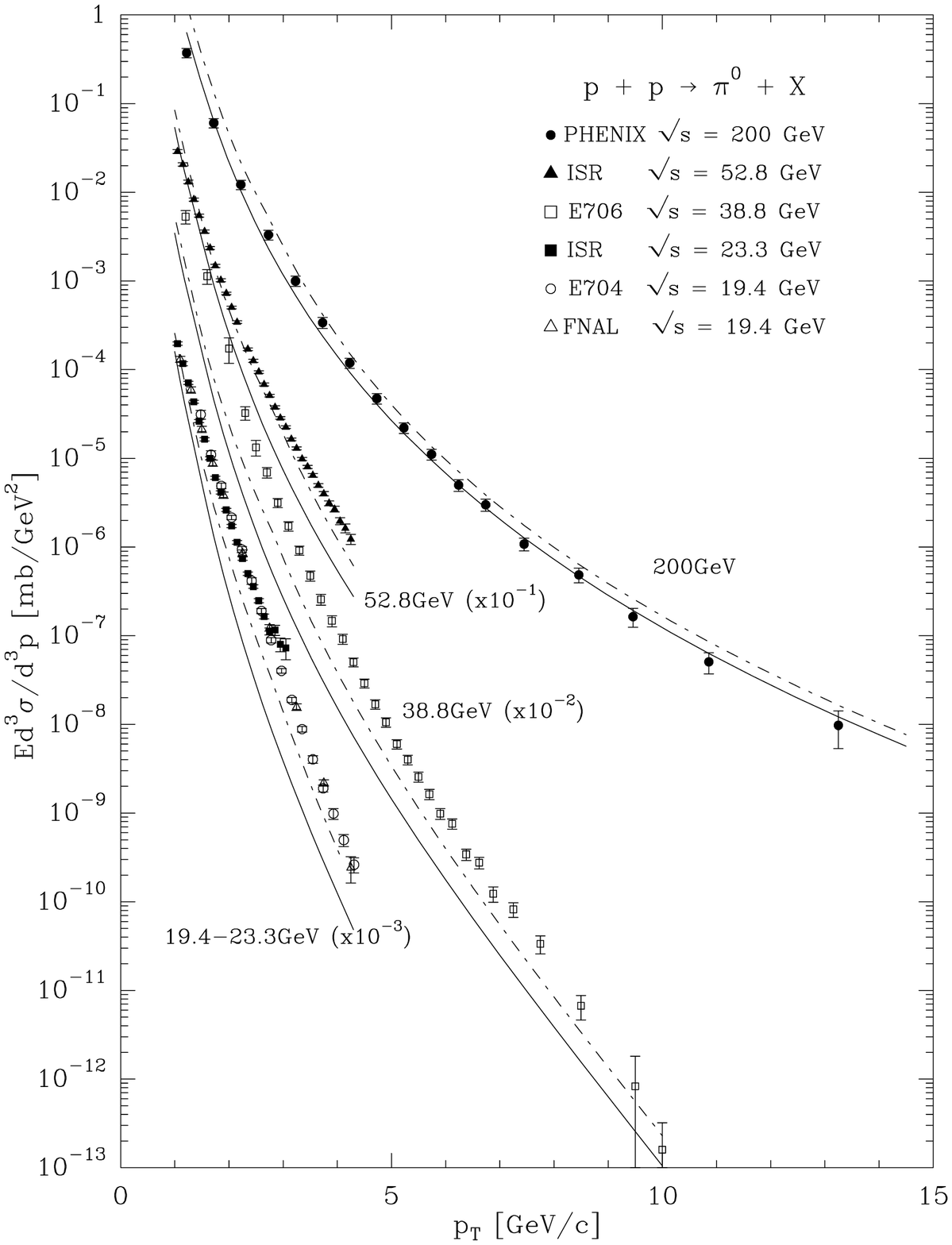,width=7cm,height=8cm}
\end{minipage}
\begin{minipage}[t]{.4\linewidth}
\epsfig{file=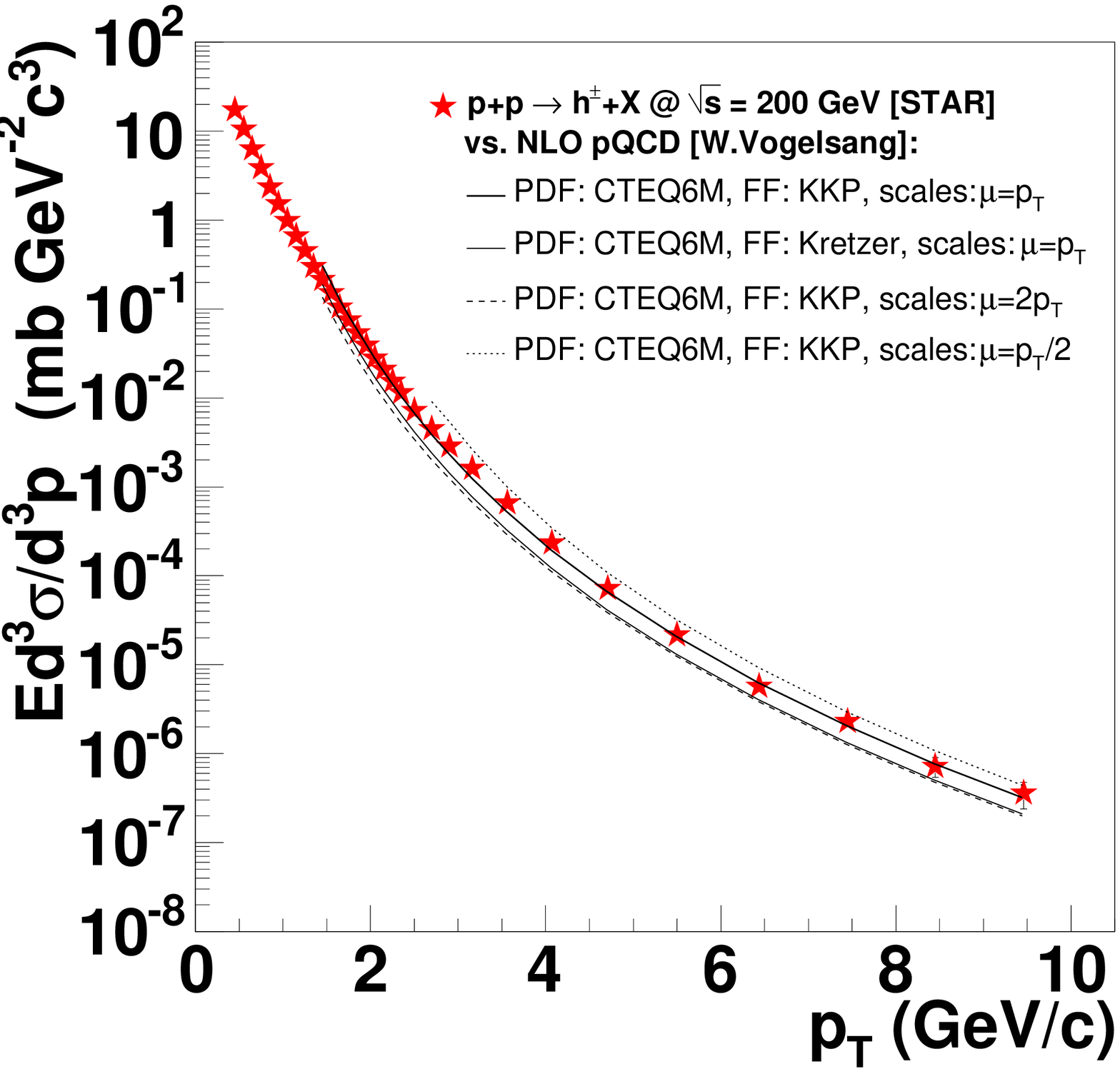,width=6.8cm,height=7.7cm}
\end{minipage}
\end{center}
\vspace{-0.5cm}
\caption{Invariant cross-sections as a function of $p_T$ 
measured at midrapidity in $p+p$ collisions at $\sqrt{s}$ = 200 GeV
compared to NLO pQCD calculations (with scales $\mu = p_T$, solid lines, 
and $\mu = p_T/2$, dotted- dashed lines) for: $p+p\rightarrow\pi^0+X$ (PHENIX
data, left, compared to results at different $\sqrt{s}$~\protect\cite{bourre}), 
and $p+p\rightarrow h^\pm+X$ (STAR, right).}
\label{fig1}
\end{figure}

High $p_T$ production in proton-proton collisions provides the baseline 
``free space'' reference to which one compares heavy-ion results in order
to extract information about the QCD medium properties. 
At RHIC, the differential cross-sections for $\pi^0$~\cite{phnx_pp_pi0_200} 
and charged hadrons~\cite{star_pp_chhad_200,brahms} above 
$p_T\approx$ 2 GeV/$c$ measured in $p+p$ collisions at $\sqrt{s}$ = 200 GeV
are well reproduced by standard next-to-leading-order (NLO) pQCD calculations 
(Figure~\ref{fig1}). This is at variance with measurements at lower 
center-of-mass energies ($\sqrt{s}\lesssim$ 65 GeV, Fig.~\ref{fig1}, left) 
where the $p_T<$ 5 GeV/$c$ cross-sections in hadronic collisions at fixed-target 
and CERN-ISR collider energies are underpredicted~\cite{aurenche,bourre} by 
pQCD calculations (even supplemented with soft-gluon resummation 
corrections~\cite{resumm}), and additional non-perturbative effects 
(e.g. intrinsic $k_T$~\cite{e706_kt}) must be introduced to bring parton 
model analysis into agreement with data. Hard production in $p+p$ collisions 
in the collider regime of RHIC seems to be basically free of the 
``distortive'' non-perturbative effects that are important at lower 
energies, and constitutes thus a experimentally and theoretically well 
calibrated baseline for heavy-ion studies.

\section{Hard spectra in hot QCD matter: central $A+A$ collisions}

\subsection{QCD factorization in $A+A$ collisions}
%\label{}
Implicitly at the root of the QCD description of high $p_T$ particle 
production in hadronic collisions is the idea that the large-$Q^2$ 
scattering between the two partons from each hadron is an {\it incoherent}
process. Namely, that the characteristic time of the parton-parton 
interaction is much shorter than any long-distance interaction 
occurring before (among partons belonging to the same hadronic 
wave function) or after (during the evolution of the struck 
partons into their hadronic final-state) the hard collision itself. 
The ``factorization theorem''~\cite{factor} reflects this mutual independence 
of QCD dynamics at different time (length) scales: the inelastic cross-section 
for the production of a given hadron $h$ in a hard process, $E\,d\sigma^{hard}_{h}/d^3p$, 
is the (factorized) product of long-distance (non-perturbative but universal 
parton distribution functions, $f_{q,g}$, and fragmentation functions, 
$D_{q,g/h}$) and short-distance (perturbatively computable parton-parton 
scattering) contributions. In nucleus-nucleus reactions, QCD factorization reads:
\begin{equation}
E\frac{d\sigma^{hard}_{AB\rightarrow h}}{d^3p} = f_{a/A}(x,Q^2)\otimes f_{b/B}(x,Q^2)\otimes 
\frac{d\sigma_{ab\rightarrow c}^{hard}}{d^3p} \otimes D_{c/h}(z,Q^2).%+ \mathcal{O}(1/Q^2)
\label{eq:factor}
\end{equation}
Since partons are effectively ``frozen'' during the hard scattering,
one can treat each nucleus as a collection of free partons. Thus, 
{\it with regard to high $p_T$ production}, the density of partons in a 
nucleus with atomic number $A$ should be equivalent to the superposition 
of $A$ independent nucleons:
\begin{equation}
f_{a/A}(x,Q^2)=A\cdot f_{a/N}(x,Q^2).
\label{eq:nPDF}
\end{equation}
From (\ref{eq:factor}) and (\ref{eq:nPDF}) it is clear that QCD factorization
implies that hard inclusive cross-sections in a minimum-bias $A+B$ reaction 
scale simply as $A\cdot B$ times the corresponding $p+p$ cross-sections:
\begin{equation}
E\,d\sigma_{AB\rightarrow h}^{hard}/d^3p=A\cdot B \cdot E\,d\sigma_{pp\rightarrow h}^{hard}/d^3p .
\label{eq:AB_scaling}
\end{equation}
%Thus, in the absence of medium effects that may distort the production
%of high $p_T$ hadrons, this simple relation permits to predict the 
%expected high $p_T$ cross-sections in heavy-ion collisions from the 
%corresponding ones in $p+p$ collisions. 
Since nucleus-nucleus experiments usually measure invariant {\it yields} 
for a given centrality bin (or impact parameter $b$), one writes instead: 
\begin{equation}
E\,dN_{AB\rightarrow h}^{hard}/d^3p\,(b)=\langle T_{AB}(b)\rangle\cdot E\,d\sigma_{pp\rightarrow h}^{hard}/d^3p ,
\label{eq:TAB_scaling}
\end{equation}
where $T_{AB}(b)$ is the Glauber geometrical nuclear overlap function 
at $b$\footnote{Since the number of inelastic nucleon-nucleon collisions 
at $b$, $N_{coll}(b)$, is proportional to $T_{AB}$: 
$N_{coll}(b) = T_{AB}(b)\cdot \sigma_{pp}^{inel}$,
one also writes Eq. (\ref{eq:TAB_scaling}) as: 
$E\,dN_{AB\rightarrow h}^{hard}/d^3p\,(b)=\langle N_{coll}(b)\rangle\cdot E\,dN_{pp\rightarrow h}^{hard}/d^3p$.}.
%(``$N_{coll}$ scaling'').}. 
One can thus quantify the medium effects on the production of
a given particle at high $p_{T}$ via the {\it nuclear modification factor}:
\begin{equation} 
R_{AB}(p_{T},y)\,=\frac{\mbox{\small{``hot QCD medium''}}}{\mbox{\small{``QCD vacuum''}}}\,=\,\frac{d^2N_{AB}/dy dp_{T}}{\langle T_{AB}(b)\rangle\,\times\, d^2 \sigma_{pp}/dy dp_{T}},
\label{eq:R_AA}
\end{equation}
which measures the deviation of $A+B$ at %impact parameter 
$b$ from an incoherent superposition of nucleon-nucleon ($NN$) 
collisions, in terms of suppression ($R_{AA}<$1) or enhancement
($R_{AA}>$1).

\subsection{High $p_T$ suppression in central $A+A$: $\sqrt{s_{_{NN}}}$ and $p_T$ dependence}
\begin{figure}[htbp]
\begin{center}
\epsfig{file=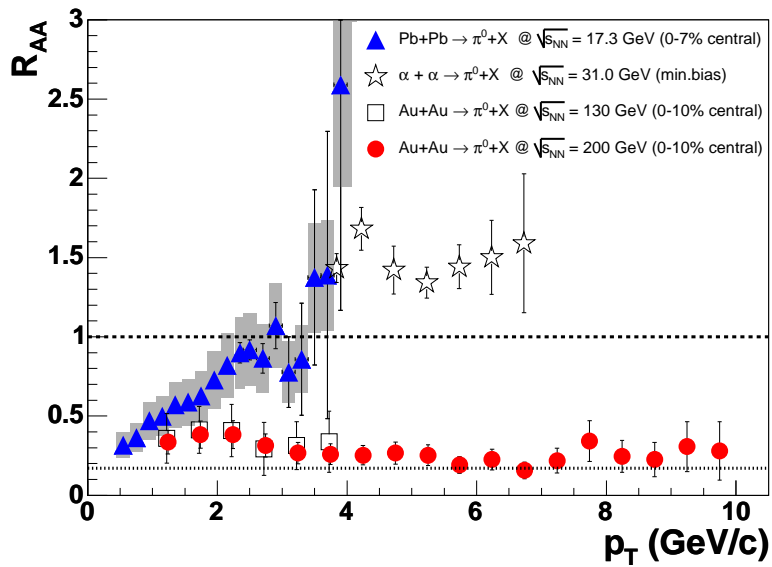,height=7.5cm}
\end{center}
\vspace{-0.5cm}
\caption{Nuclear modification factor, $R_{AA}(p_T)$, for $\pi^0$ measured in
central nucleus-nucleus reactions at SPS~\protect\cite{wa98,denterria}, 
ISR~\protect\cite{ISR_pi0}, and RHIC~\cite{ppg003,ppg014}. 
The dashed (dotted) line is the expectation of ``$N_{coll}$ ($N_{part}$) 
scaling'' for hard (soft) particle production.}
\label{fig:R_AA_pi0}
\end{figure}

%\label{}
One of the most interesting results at RHIC so far is the 
{\it breakdown} of the expected incoherent parton scattering assumption 
for high $p_{T}$ production, Eq. (\ref{eq:TAB_scaling}), observed in 
central $Au+Au$. Figure~\ref{fig:R_AA_pi0} shows $R_{AA}$ as a function 
of $p_T$ for $\pi^0$ produced in nucleus-nucleus reactions at 
different center-of-mass energies. RHIC data at 200 GeV (circles) 
and 130 GeV (squares)~\cite{ppg003,ppg014} are noticeably below unity 
in contrast to the enhanced production observed in $\alpha+\alpha$ 
collisions at CERN-ISR~\cite{ISR_pi0} (stars). 
This enhanced production, observed first in $p+A$ fixed-target 
experiments (``Cronin effect'')~\cite{cronin}, is interpreted in terms 
of multiple initial-state soft and semi-hard interactions which 
broaden the transverse momentum of the colliding partons prior to 
the hard scattering.
The situation at CERN-SPS is not completely clear. 
Whereas the original work~\cite{wa98} reported a strong Cronin 
enhancement, a recent reanalysis using a better $p+p \rightarrow \pi^0+X$
reference~\cite{denterria} shows that the 0--7\% most central $Pb+Pb$ 
data is consistent with $R_{AA}\approx$ 1 (triangles in Fig.~\ref{fig:R_AA_pi0}) 
and that the production in head-on $Pb+Pb$ reactions (0--1\% central) is 
actually suppressed ($R_{AA}\approx$ 0.6) indicating that some amount of 
``jet quenching'' may already be present %in central heavy-ion reactions 
at $\sqrt{s_{\mbox{\tiny{\it{NN}}}}}\approx$ 20 GeV.
A concurrent measurement at RHIC of high $p_T$ hadron spectra in 
$Au+Au$ {\it and} $p+p$ collisions at these lower $\sqrt{s}$ would definitely 
set the issue of the onset of the suppression in central $A+A$ reactions.

The breakdown of the expectations from collinear factorization for 
high $p_T$ production in central $Au+Au$ collisions at RHIC, has been 
interpreted as indicative of:
\begin{enumerate}
\item Strong {\bf initial-state} effects: The parton distribution 
functions in the nuclei are strongly modified: 
$f_{a/A}<<A\,\cdot f_{a/p}$ in the relevant ($x,Q^2$) range, 
resulting in an effective reduction of the number of 
partonic scattering centers in the initial-state.
\item Strong {\bf final-state} effects: The parton fragmentation 
functions (or, more generally, any post hard collision effect 
on the scattered partons) are strongly modified in the 
nuclear medium compared to free space.
\end{enumerate}
Explanation (i) is usually invoked in the context of the 
``Color-Glass-Condensate'' picture~\cite{cgc} which assumes 
that the kinematical conditions prevailing in the initial-state of 
an atomic nucleus boosted to RHIC energies are such that nonlinear 
QCD effects ($g+g\rightarrow g$ processes, amplified by a $A^{1/3}$ 
factor compared to the proton case) are important and lead to a  
saturation of the strongly rising small-$x$ gluon densities in the nuclei. 
Leading-twist QCD factorization itself breaks down since the incoherence 
between long- and short-distance effects in which the product 
Eq. (\ref{eq:factor}) relies upon, does not hold anymore.
CGC calculations predict $N_{part}$ (instead of $N_{coll}$) 
scaling at moderately high $p_T$'s, as approximately 
observed in the data (dotted line in Fig.~\ref{fig:R_AA_pi0}).
Explanation (ii) relies on the expectations of ``jet quenching''~\cite{vitev}
in a Quark Gluon Plasma in which the hard scattered partons lose energy 
by final-state ``gluonstrahlung'' in the dense partonic system formed 
in the reaction. After traversing the medium, the partons fragment into 
high $p_T$ (leading) hadrons with a reduced energy compared to 
standard fragmentation in the ``vacuum''. Different jet quenching 
calculations can reproduce the magnitude of the $\pi^0$ suppression 
assuming the formation of a hot and dense system characterized 
by different, but closely related, properties~\cite{vitev}: 
i) large initial gluon densities $dN^{g}/dy\approx$ 1100, %~\cite{vitev}, 
ii) large ``transport coefficients'' $\hat{q}_{0}\approx$~3.5~GeV/fm$^2$, %~\cite{arleo}, 
iii) high opacities $L/\lambda\approx$ 3.5, %~\cite{levai_pp_pi0}, 
iv) effective parton energy losses of the order of $dE/dx\approx$~14 GeV/fm, %~\cite{xnwang}, 
or v) plasma temperatures of $T\approx$ 0.4 GeV.%~\cite{moore}.

\subsection{High $p_T$ suppression in central $A+A$: particle species dependence}
%\label{}

\begin{figure}[htbp]
%\hspace*{-1.8cm}
\begin{center}
\epsfig{file=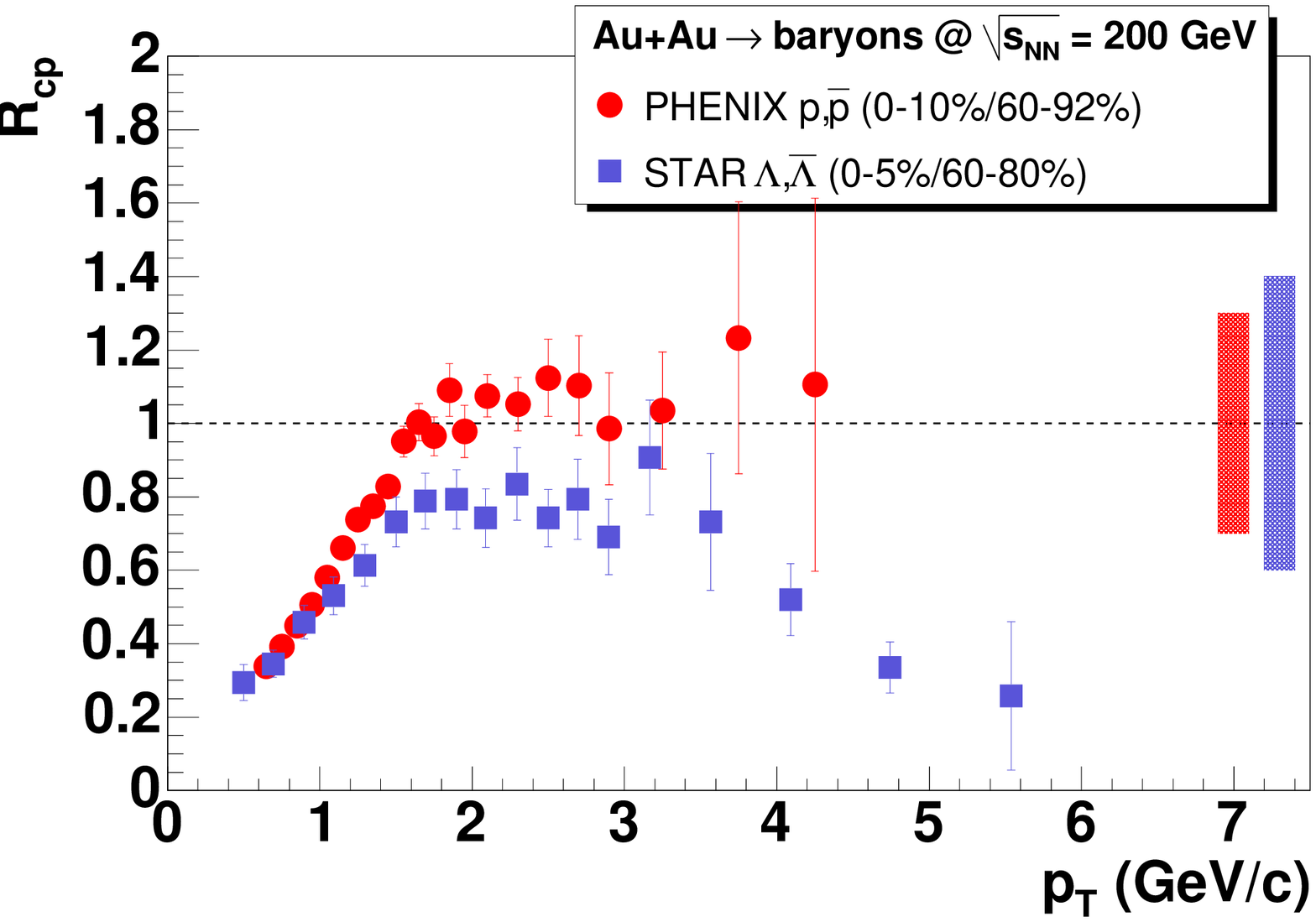,height=5.2cm}
%\hspace*{4mm}
\epsfig{file=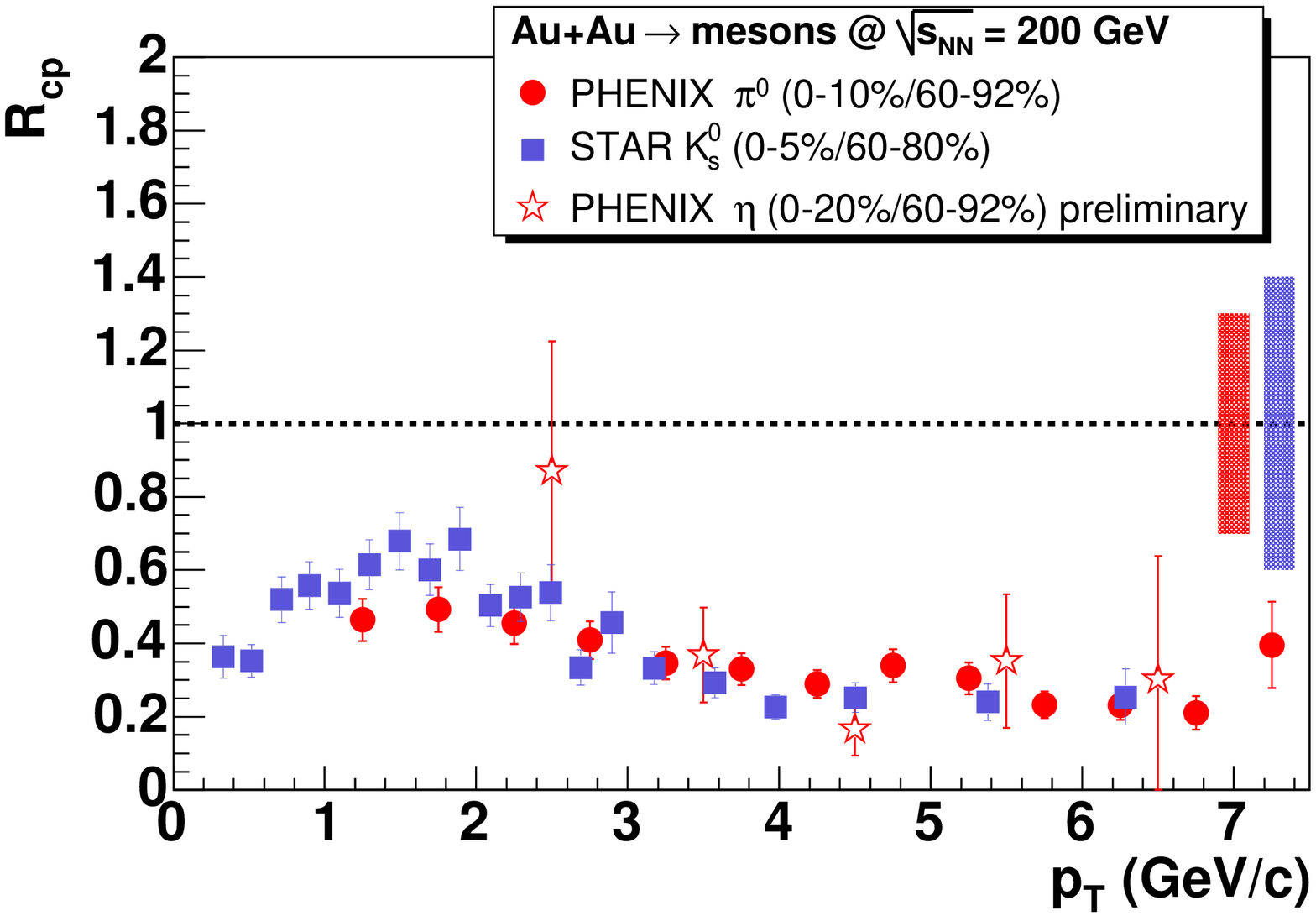,height=5.2cm}
\end{center}
\vspace{-0.5cm}
\caption[]{Ratio of central over peripheral $N_{coll}$ scaled yields, $R_{cp}$, 
as a function of $p_{T}$ for baryons (left): $(p+\bar{p})/2$ (dots) and 
$(\Lambda+\bar{\Lambda})/2$ (squares); and for mesons (right): $\pi^0$ (dots), 
$K^0_s$ (squares) and $\eta$ (stars); measured in $Au+Au$ collisions at 200 GeV.
The shaded bands indicate the associated fractional normalization uncertainties.}
\label{fig:flavor_dep}
\end{figure}

Another intriguing result of the RHIC program is the different suppression 
pattern of baryons and mesons at moderately high $p_{T}$.
Figure~\ref{fig:flavor_dep} shows the $N_{coll}$ scaled central 
to peripheral yield ratios\footnote{Since the peripheral $Au+Au$ 
(inclusive and identified) spectra scale with $N_{coll}$ when compared 
to the $p+p$ yields, the ratio $R_{cp}$ carries basically the same 
information as the nuclear modif. factor $R_{AA}$.}, 
$R_{cp}$, for baryons (left) and mesons (right). In the range 
$p_T\approx$ 2 -- 4 GeV/$c$ the (anti)protons are not suppressed 
($R_{cp}\sim$ 1) at variance with the pions which are reduced 
by a factor of 2 -- 3. The resulting baryon/meson$\sim$0.8 ratio %in this $p_T$ range 
is clearly at odds with the ``perturbative'' $\sim$0.2 
ratio measured in $p+p$ or $e^{+}e^{-}$ collisions. Such a 
particle composition is inconsistent with standard fragmentation 
functions, and points to an additional non-perturbative mechanism for 
baryon production in central $Au+Au$ reactions in this intermediate $p_T$ range. 
In the recombination picture~\cite{fries}, the quarks present in the dense
environment coalesce and with the addition of their momenta, the 
soft production of baryons extends to larger $p_{T}$ values than that 
for mesons. Beyond $p_{T} \approx 5$ GeV/$c$ fragmentation becomes the 
dominant production mechanism for all species. Estimates of the formation time 
of the (leading) hadrons qualitatively support this scenario. A hard scattered 
parton with momentum $p$ = 3 (10) GeV/$c$ hadronizes into a fully formed meson 
of radius $R_h\approx$ 0.8 fm in a time~\cite{dokshitzer} 
$\tau_h\approx p\cdot R_h^2\approx$ 10 (30) fm/$c$.
The total lifetime of the strongly interacting system produced in
$Au+Au$ reactions is $\tau\approx$ 15 fm/$c$ as extracted from different
experimental observables~\cite{fabrice}. Thus, partons with moderate 
energies (leading to hadrons with $p_T\approx$ 2 -- 4 GeV/$c$) will 
not fragment in the vacuum, as do the more energetic ones, but in a 
environment where they can still recombine with other surrounding 
particles. %di-quarks ?

%\subsection{Hard production in central $A+A$: heavy-quarks, direct photons}
%%\label{}
%Electrogmanetic probes insensitive to colored final-state do show
%``collision scaling'' at high pT

\section{Hard spectra in cold QCD matter: $d+A$ and $l+A$ collisions}

%In lepton- and proton- (or deuteron-) nucleus collisions,
% at variance with nucleus-nucleus reactions,
%no dense and hot QCD medium is expected to be formed in the final state. 
Hard scattering in lepton- and proton- (or deuteron-) nucleus 
collisions allows the study of the properties of the 
nuclear wave-function with minimal final-state distortions due to 
{\it dense} QCD medium effects. In 2003, RHIC run $d+Au$ collisions
as a ``control'' experiment in order to disentangle 
between the two different scenarios (QGP and CGC) proposed 
to explain the high $p_T$ deficit observed in central $Au+Au$ 
at mid-rapidity.

\subsection{High $p_T$ production at midrapidity: Cronin enhancement}
\begin{figure}[htbp]
\begin{center}
\epsfig{file=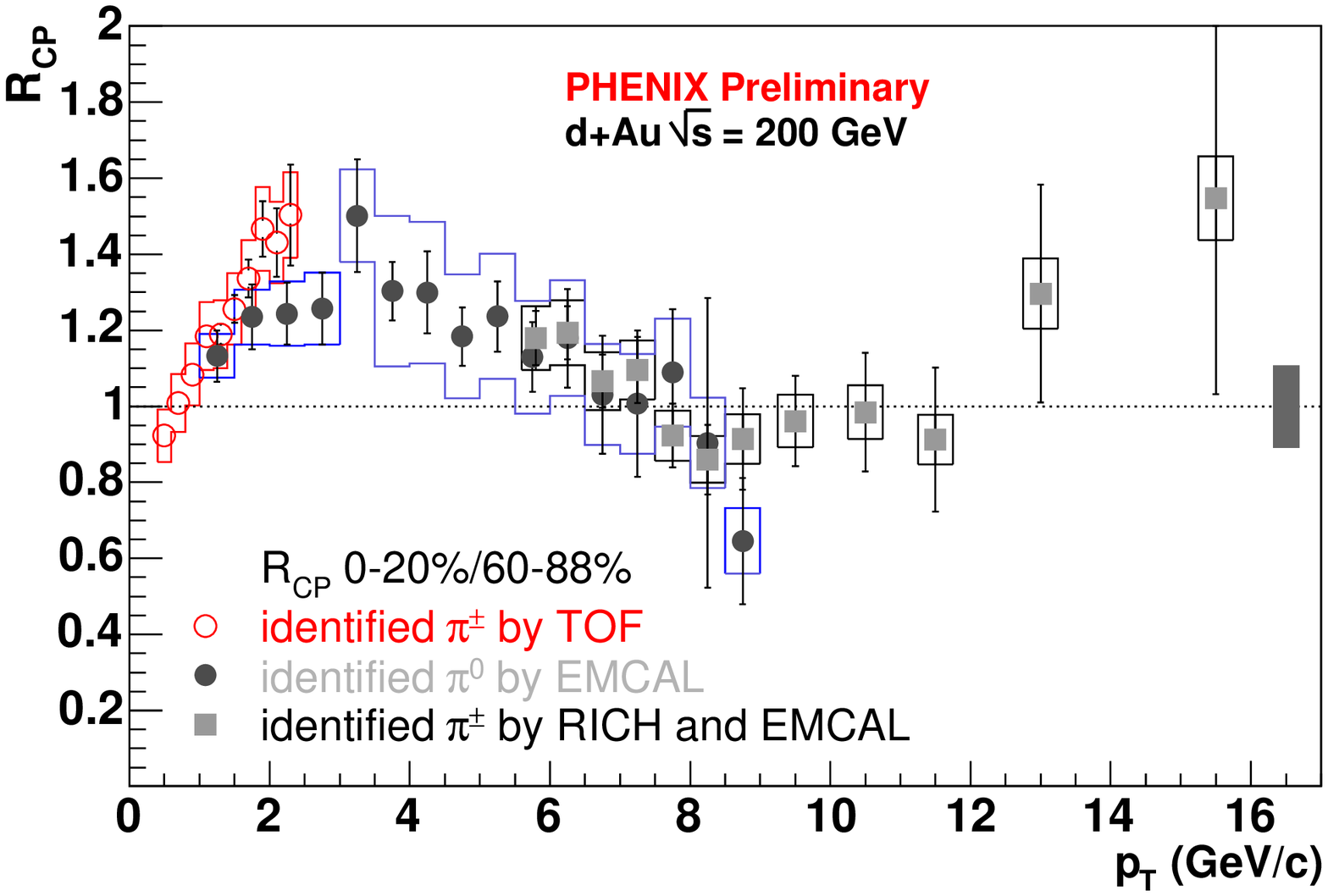,height=5.8cm,width=8.5cm}
\epsfig{file=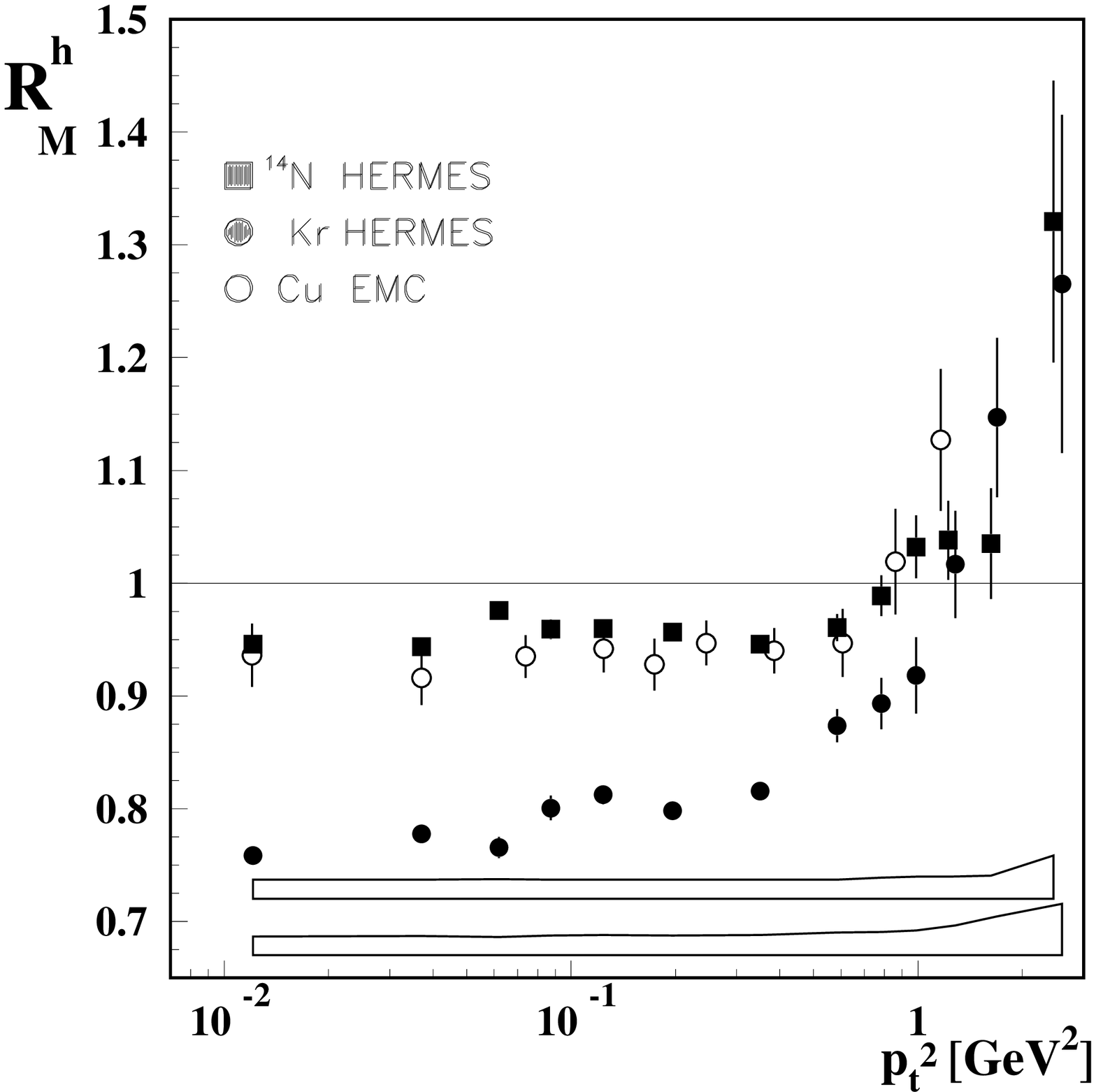,height=5.8cm,width=7cm}
\end{center}
\vspace{-0.5cm}
\caption{Left: Ratio of central over peripheral $N_{coll}$ scaled 
yields, $R_{cp}$, as a function of $p_{T}$ for pions measured by PHENIX at $y$ = 0 
in $d+Au$ at $\sqrt{s_{_{NN}}}$ = 200~GeV. Right: Nuclear attenuation
factor, $R_M^h$, versus $p_T^2$ for charged hadrons measured by HERMES 
in $e+N,Kr$ at $\sqrt{s}$ = 27.6 GeV and by EMC in $\mu+Cu$ collisions
%(HERMES, closed symbols) compared to $\mu$+Cu data (EMC, open circles)
~\protect\cite{hermes}.}
%Multiplicity ratio for charged hadrons versus pt2 for small nu, Greek>7 GeV 
%and z>0.2. The HERMES data on Kr and 14N are compared to the EMC [18] data 
%for Cu in the range 10<small nu, Greek<80 GeV. The error bars represent the 
%statistical uncertainties. The systematic uncertainty for Kr (14N) is shown as the lower (upper) band. 
\label{fig:cronin_coldA}
\end{figure}

%\label{}
The results of high $p_T$ pion production at y = 0 in $d+Au$ 
collisions at $\sqrt{s_{_{NN}}}$ = 200~GeV do not show any 
indication of suppression (Fig.~\ref{fig:cronin_coldA}, left). 
On the contrary, pion yields appear Cronin enhanced 
($R_{cp}>$ 1) compared to the expectations of collinear 
factorization\footnote{Note in Fig.~\ref{fig:cronin_coldA} left, 
that the Cronin effect seems to disappear ($R_{cp}\approx$ 1) above 
$p_T\approx$ 8 GeV/$c$, a result found also in $p+A$ collisions 
at fixed-target energies~\cite{cronin}.}. 
This result indicates, in a model-independent way, that the 
observed suppression in $Au+Au$ central collisions is not 
an initial-state effect arising from strong modifications 
of the gluon distribution function in nuclei as proposed by 
CGC approaches, %at moderately small values of parton fractional momenta $x$, 
but results instead from a final-state effect in the produced 
dense medium. 
The Cronin enhancement is also observed above $p_T^2\approx$ 1 GeV$^2/c^2$ 
in the HERMES and EMC deep-inelastic ``hadron multiplicity ratio'' $R_M^h$
(ratio of the number of hadrons of type $h$ produced per DIS event on a 
nuclear target to that from a deuterium target) for $^{14}N$, $Cu$ and $Kr$ 
nuclei~\protect\cite{hermes} (Fig. \ref{fig:cronin_coldA}, right). 
Two points are worth noticing here: (i) at such relatively high values of 
$p_T^2$, the exchanged virtual photon interacts directly with the 
{\it partonic} constituents of the nucleus, and (ii) any $p_T$ 
broadening is due to multiple scattering of the {\it outgoing} quark 
propagating inside the (cold) nuclear medium. Thus, the observed 
Cronin effect in lepton-nucleus collisions is 
due to {\it final-state partonic} multiple scattering.

\subsection{High $p_T$ production at forward rapidities: searching for gluon saturation}
\begin{figure}%[htbp]
\begin{center}
\epsfig{file=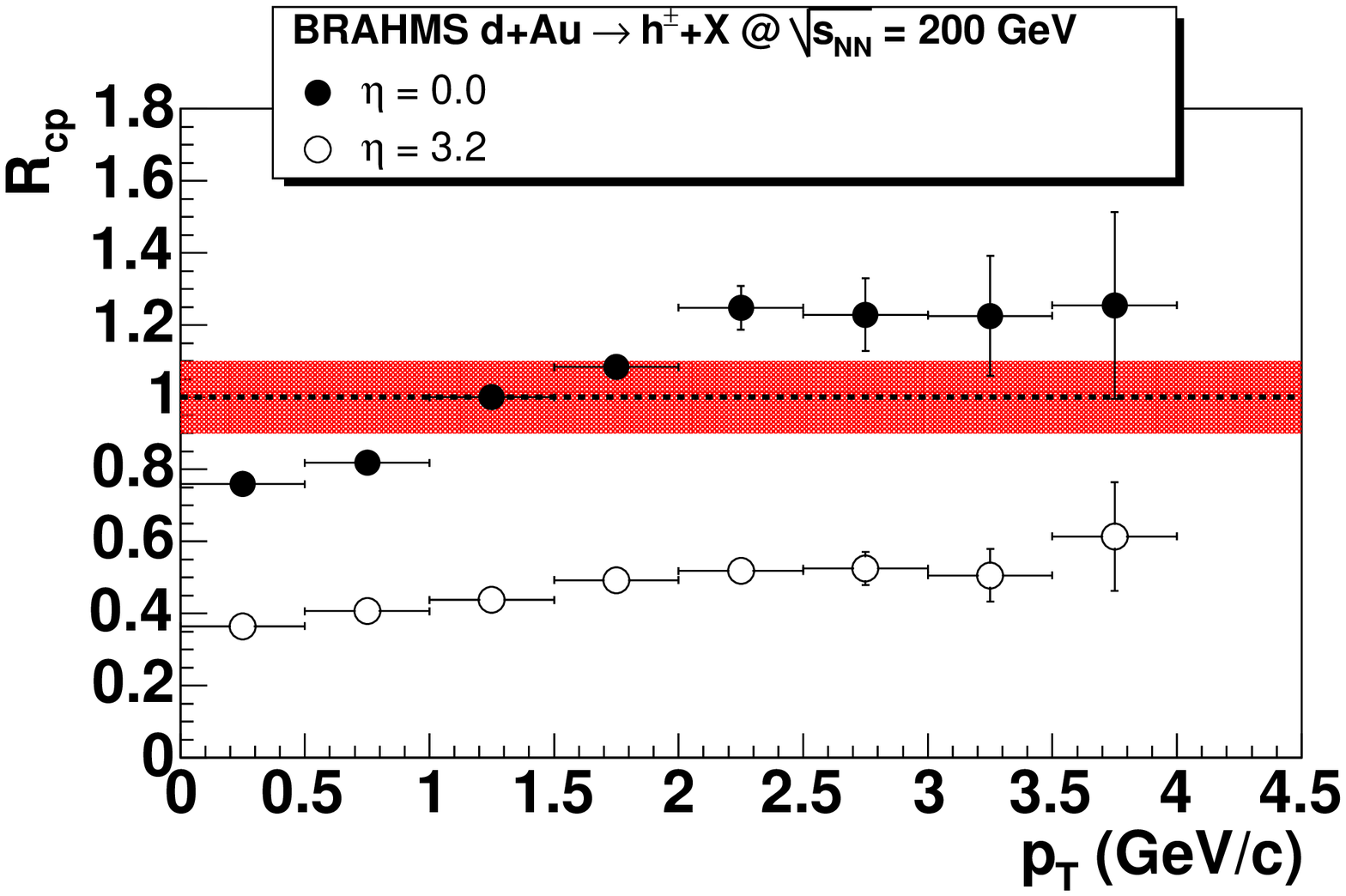,width=7.5cm,height=5.5cm}
\hspace{0.5cm}
\epsfig{file=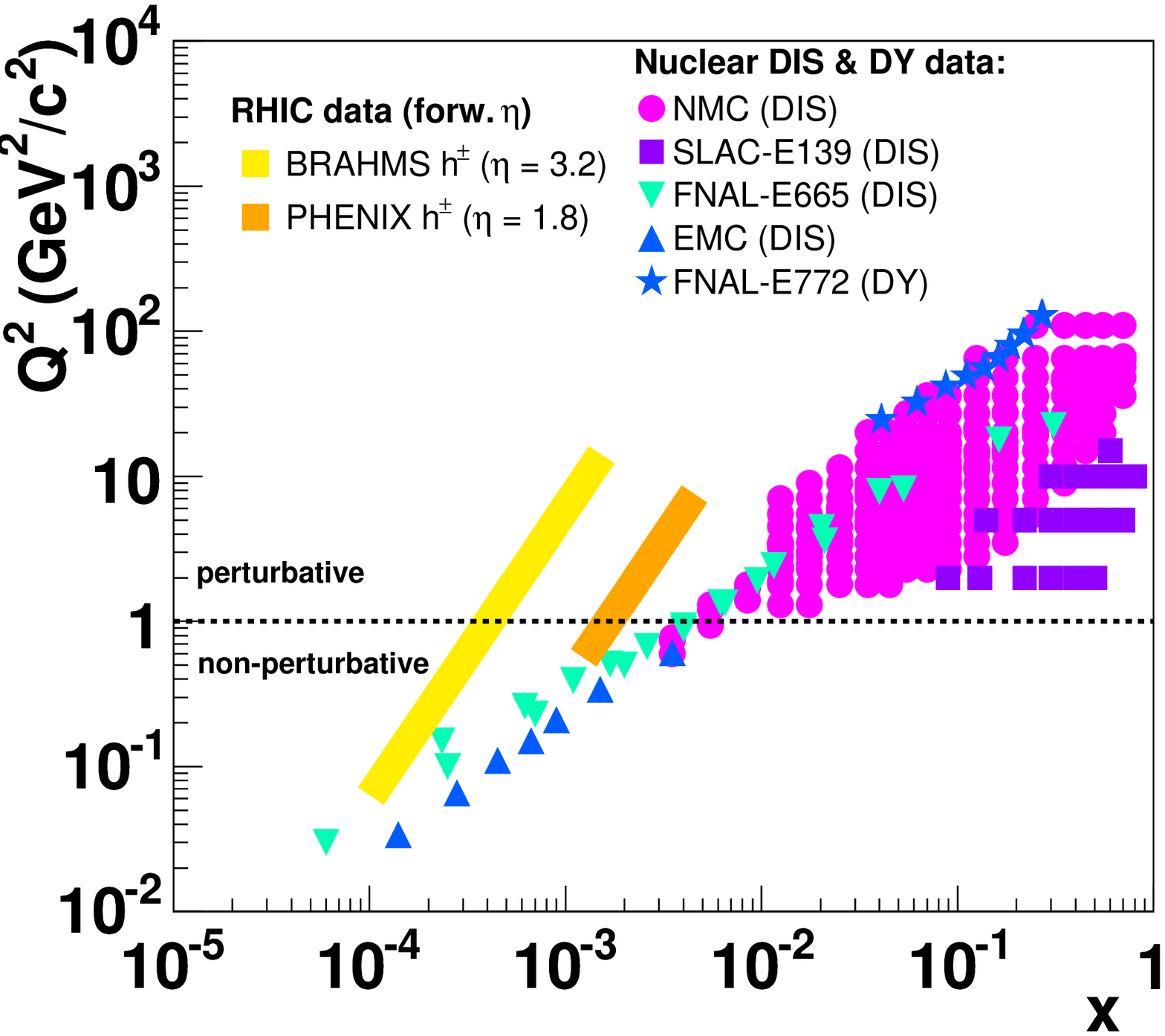 ,width=6.8cm,height=6.0cm}
\end{center}
\vspace{-0.5cm}
\caption{Left: Ratio of central over peripheral $N_{coll}$ scaled 
yields, $R_{cp}$, as a function of $p_{T}$ for charged hadrons measured by 
BRAHMS at $\eta$ = 0 (dots) and $\eta$ = 3.2 (open circles) 
in $d+Au$ at $\sqrt{s_{_{NN}}}$ = 200~GeV~\protect\cite{brahms}. 
Right: Kinematical range in the $x$-$Q^2$ plane probed in nuclear DIS
and DY processes, and in $d+Au$ at forward rapidities at RHIC.}
\label{fig:Rcp_brahms}
\end{figure}

%\label{}
In a sense, the apparent absence of gluon saturation effects in 
hard $d+Au$ cross-sections at $y$ = 0 at RHIC is not completely 
surprising inasmuch as the kinematical range probed corresponds to 
relatively moderate values of Bjorken $x\approx 2p_T/\sqrt{s}\approx$ 10$^{-2}$
where standard DGLAP evolution describes well the DIS data at HERA.
A simple way to probe smaller values of $x_2$ in the $Au$ nucleus 
consists in looking at hadron production in the forward direction. 
Since %making use of the $e^{-y}$ kinematical factor 
$x_{1,2} = p_T/\sqrt{s}(e^{\pm y_1}+e^{\pm y_2})$ for a $2\rightarrow 2$ process, 
$x$ decreases by a factor of $\sim$10 for every 2-units of rapidity 
one moves away from $y$ = 0. 
BRAMHS~\cite{brahms} and PHENIX~\cite{mliu} results on high $p_T$ 
charged hadron production at pseudorapidities $\eta$ = 3.2 and 
$\eta$ = 1.8 (corresponding to $x\approx$ $\mathcal{O}$(10$^{-4}$) 
and $\mathcal{O}$(10$^{-3}$) respectively) show a suppression instead of 
an enhancement as found at $\eta$ = 0 (Fig.~\ref{fig:Rcp_brahms}, left). 
Interestingly, this is the first time that the nuclear PDFs are probed 
at such small values of $x$ in the {\it perturbative} domain 
($Q^2\approx p_T^2>$ 1 GeV$^2/c^2$) (Fig.~\ref{fig:Rcp_brahms}
right). BRAHMS $R_{cp}\approx$ 0.5 result seems to indicate
that the ratio of $Au$ over $p$ gluon densities is
$R_{G}^{Au}(x\approx 10^{-4},Q^2\approx$ 2 GeV$^2/c^2)\approx$ 0.5, 
whereas standard {\it leading-twist} DGLAP analysis of the 
nuclear PDFs (based on global fits of the DIS and 
Drell-Yan data above $Q^2$ = 1 GeV$^2/c^2$ shown in 
Fig.~\ref{fig:Rcp_brahms}, right) indicate a less significant 
amount of gluon ``shadowing'' in this range: 
$R_{G}^{Au}\approx$ %$R_{G}^{Au}(x\approx 10^{-4},Q^2\approx$ 2 GeV$^2/c^2)\approx$
0.8~\cite{eks98,deflorian}. Whether this larger suppression is 
due to soft physics (the global $dN/dy$ distributions in 
lower energy $p+A$ collisions are also found to be depleted at 
forward $\eta$~\cite{rachid}) or a genuine CGC effect, is still matter of 
discussion at this point.

\section{Summary}
During its first four years of operation, RHIC has provided 
%opened up a new window to many-body QCD at high energies
many new and exciting results on the %behaviour of the 
many-body dynamics of QCD at high energies. The suppressed high $p_T$ 
hadroproduction observed in central $Au+Au$ reactions and 
in $d+Au$ collisions at forward-rapidities is inconsistent 
with the basic QCD factorization expectations that describe 
particle production in $p+p$ at $\sqrt{s}$ = 200~GeV. 
The factor of 4--5 suppression in central $Au+Au$ is unambiguously 
due to final-state effects (since no such an effect is seen
in $d+Au$ collisions at $y$ = 0) and can be reproduced by 
calculations of parton energy loss in a strongly interacting 
medium with energy densities well above those where lattice 
QCD predicts a transition to a Quark Gluon Plasma. The factor 
of $\sim$2 deficit observed at $y\approx$ 3 in $d+Au$ reactions 
may be the first empirical indication of higher-twist 
(non-linear) QCD effects at small Bjorken-$x$ 
in the hadronic wave functions.

%\section{Acknowledgments}
%I would like to thank D.~de~Florian, C.~Salgado, I.~Vitev and W.~Vogelsang for 
%useful discussions.%and for providing me with different results that 
%%helped produce some of the plots presented here.

\end{document}